\documentclass[11pt, letterpaper, twocolumn, journal]{IEEEtran}

\IEEEoverridecommandlockouts

\usepackage[cmex10]{amsmath}
\usepackage{amssymb}
\usepackage{subfigure}
\usepackage{graphicx,epsfig}
\usepackage{cite}
\usepackage{url}
\usepackage[usenames]{color}

\begin{document}

\title{A Tight Upper Bound on Bit Error Rate of Joint OFDM and Multi-Carrier Index Keying}

\author{
Youngwook~Ko,~\IEEEmembership{Member,~IEEE,}
%and~Author~2,~\IEEEmembership{Member,~IEEE}
        % stops a space
\thanks{Youngwook~Ko is with the Electronics, Communications and Information Technology Institute, Queen's University Belfast, Belfast, BT3 9DT United Kingdom, e-mail: y.ko@qub.ac.uk.}
}

\maketitle

\vspace{-1.9cm}
\begin{abstract}
This letter investigates the performance enhancement by the concept of multi-carrier index keying in orthogonal frequency division multiplexing (OFDM) systems.
For the performance evaluation, a tight closed-form approximation of the bit error rate (BER) is derived introducing the expression for the number of bit errors occurring in both the index domain and the complex domain, in the presence of both imperfect and perfect detection of active multi-carrier indices. The accuracy of the derived BER results for various cases are validated using simulations, which can provide the accuracy within 1~dB at favorable channels.
\end{abstract}

%\vspace{-0.5cm}

%\thispagestyle{empty}
%\clearpage
%\setcounter{page}{1}

\section{Introduction}
\label{sec:intro}

Orthogonal frequency division multiplexing (OFDM) has been adopted in the majority of today and future communication standards such as IEEE 802.11, 3GPP's LTE-Advanced, due to its robustness to multipath fading. The performance of these systems with increased sub-carriers is heavily dependent on an increased sensitivity to mismatched conditions such as frequency offset and phase noise \cite{pollet_tcom95} as well as transmission nonlinearity caused by the non-constant power ratio of OFDM symbols \cite{goeckel_ofdm02,baxley_ofdm04}.

In \cite{haas_pimrc09,haas_globecom11}, the so-called sub-carrier index modulation scheme modified the classical OFDM systems treating the sub-carrier index as additional resource to decrease the bit error rate (BER) faster than the classical OFDM at a low complexity with only a few sub-carrier activation. The effects of channel estimation errors on the approximate pairwise error probability (PEP) of the OFDM modulating the index of sub-carrier was more recently discussed in \cite{basar_tsp13}.

The contribution of this letter is twofold. We first generalizes the BER expression of a joint multi-carrier index keying and OFDM (MCIK-OFDM) that is based on any number of active sub-carriers and includes expressions for the number of bit errors by both the MCIK and the OFDM transmissions.
In \cite{haas_pimrc09,haas_globecom11}, the BER is limited by a fixed number of active sub-carriers that differs from what we consider herein. For example, the approach in \cite{haas_globecom11} cannot be used directly with both a small and a large number of active sub-carriers. Our contribution is secondly to analyze the performance of the MCIK-OFDM system deriving a tight upper bound on the BER in the presence of imperfect and perfect detection of active indices.

\section{Joint MCIK-OFDM System Model}
\label{sec:sys}

We consider a peer-to-peer M-QAM OFDM transmission with $N_c$ sub-carriers that consists of $n$ clusters of $N$ sub-carriers (i.e., $N_c=n N$).
A stream of M-QAM symbols is first serial-to-parallel converted, where every $n\, (\leq N_c)$ symbols are grouped into a vector $\mathbf{s}=[s_{1}, s_{2},\cdots,s_{n}]^T$ and $s_i\in{\cal S}$ are used to modulate sub-carriers, as in the classical OFDM, but it differs from that the modulated sub-carriers are only those of $n$ activated indices, similar to \cite{haas_pimrc09,haas_globecom11}. For the $n$ active indices, a different stream of $m_0$ bits per cluster is used to randomly select one out of $N$ indices of sub-carriers, and thus $n$ randomly activated sub-carriers at every transmit interval. In this process, namely multi-carrier index keying (MCIK),
the $n m_0$ bits streams modulate a combination of the $n$ indices of sub-carriers that are mutually modulated by the above $n$ symbols streams.
Note that there are $L=N^n$ combinations available, where for the simplicity in analysis $L$ is assumed to be $L=2^{\lfloor \log_2 {\rm B}(N_c,n) \rfloor}$ and ${\rm B}(,)$ denotes the binomial coefficient. After modulating both the active indices of sub-carriers (by MCIK) and the sub-carriers of the active indices (by OFDM), $\mathbf{s}$ is mapped to $n$ sub-carriers of the active indices. A combination of the active indices is denoted by $\mathbf{x}_l$, i.e., $\mathbf{x}_l=[i_1,\cdots,i_n]$ where $i_\beta\in\{1,\cdots,N_c\}$ for $\beta=1,\cdots,n$. Note that $N_c-n$ inactive sub-carriers are zero padded to represent no transmission of M-QAM symbols on these \cite{haas_globecom11}. Taking into account both $\mathbf{x}_l$ and $\mathbf{s}$, then, the OFDM block to transmit forms the $N_c\times 1$ OFDM block as
$\mathbf{s}_F=[s(1),\cdots,s(N_c)]^T$ where $s(k)\in\{0,{\cal S}\}, \forall k$. Unlike the classical OFDM, $\mathbf{s}_F$ in the proposed system comprises $N_c-n$ zero elements whose indices help carry information of $n m_0$ bits. Supposing that the channel has a discrete-time impulse response during the OFDM block interval in the frequency domain defined as $\mathbf{h}_F=[h(1),\cdots, h(N_c)]$ where $h(k)$ for $k\in \mathbf{x}_l$ represent Rayleigh fading channel, being independent identically distributed (i.i.d.) complex Gaussian with zero mean and unit variance, i.e., $h(k)\sim{\cal CN}(0,1)$, and others for $k\notin \mathbf{x}_l$ are zeros. The fading channel is assumed to be quasi-static so that the channel gains vary from one OFDM block to another. The input-output model in the frequency domain can be equivalently given by
\begin{equation}
\mathbf{y} = \mathbf{h} \mathbf{S} + \mathbf{n}
\label{eq:io}
\end{equation}
where $\mathbf{y}=[y(1),\cdots,y(N_c)]$, $\mathbf{h}=[h(i_1),\cdots,h(i_n)]$ with $i_\beta\in\mathbf{x}_l$, $\mathbf{S}$ is the $n\times N_c$ matrix such that
\begin{equation}
%\mathbf{S}=\begin{bmatrix}
%             \mathbf{s}_{1,\alpha_1} &  & \mathbf{0} \\
%              & \ddots &  \\
%             \mathbf{0} &  & \mathbf{s}_{n,\alpha_n} \\
%           \end{bmatrix}
\mathbf{S}=\text{diag} \left( \mathbf{s}_{1,\alpha_1}, \cdots, \mathbf{s}_{n,\alpha_n} \right)
\end{equation}
where $\mathbf{s}_{\beta,\alpha_\beta}=[\mathbf{0}_{1\times \alpha_\beta-1}, \, s(i_\beta), \,\mathbf{0}_{1\times N-\alpha_\beta}]$, $\alpha_\beta$ indicates the location of the active sub-carrier within cluster~$\beta$, i.e., $i_\beta=(\beta-1)N+\alpha_\beta$ and $\alpha_\beta\in\{1,\cdots,N\}$, $s(i_\beta) \in {\cal S}$ is the M-ary symbol, and $\mathbf{n}=[n(1),\cdots,n(N_c)]$ denotes the independent, additive white Gaussian noise (AWGN) vector, i.e., $n(k)\sim{\cal CN}(0,N_o), \forall k$. We denote the signal-to-noise ratio (SNR) by $\rho=E_s/N_o$ where $E_s$ denotes the average power for the M-QAM symbol.

\section{BER analysis}
\label{sec:ber}

The joint MCIK-OFDM scheme can transmit the total number of $m_t$ bits that is the sum of two information rates: $m_{0}=\log_2(L)=n\log_2 N$ bits by the MCIK; and simultaneously $m_1$ bits (or $n$ symbols in $\mathbf{s}$) by the OFDM, i.e., $m_1 = n\log_2 M $. Therefore, the MCIK-OFDM scheme offers $m_t=m_0 + m_1$ (bits/symbol).

Let us define the BER of the proposed scheme as the ratio of the number of bits in error to the total number of bits in transmission, which can be given:
\begin{equation}
P_b = \frac{\text{Num. of bits in error}~(m_e)}{\text{Num. of bits in transmission}~(m_t)}
\label{eq:pb}
\end{equation}
where the numerator $m_e$ is the sum of bit errors among the $n$ clusters, i.e., $m_e=\sum_{\beta=1}^n m_{e,\beta}$.

The bit errors of $m_{e,\beta}$ % at cluster~$\beta$
result from three error cases:
 \begin{itemize}
 \item[(i)] an incorrect index of active sub-carrier and an incorrect M-ary symbol;
 \item[(ii)] an incorrect index of active sub-carrier and a correct M-ary symbol; and
 \item[(iii)] a correct index of active sub-carrier and an incorrect M-ary symbol.
 \end{itemize}
The bit errors caused by only the incorrect index in cases~(i), (ii) is denoted by $m_{e_0,\beta}$ while those by the incorrect M-ary symbol in cases~(i), (iii) by $m_{e_1,\beta}$.
Thus, per cluster, we have $m_{e,\beta}=m_{e0,\beta} + m_{e1,\beta}$ with \emph{$m_{e0,\beta}$ and $m_{e1,\beta}$ being the numbers of bit errors by the MCIK and the OFDM,} respectively.

To compute \eqref{eq:pb}, we first focus on $m_{e,\beta}$ per cluster since each cluster independently modulates and demodulates $m_t/n$ bits; we examine the expressions for both $m_{e0,\beta}$ and $m_{e1,\beta}$ which has been overlooked by others in this field.
$m_{e,\beta}$ of all the clusters will be used later to express $m_e$ and thus, the overall BER.

Unlike the classical OFDM, the maximum likelihood (ML) detector only on $s(k)$ based on $y(k)$ is not sufficient to retrieve $m$ bits in the proposed system. This is because the MCIK-OFDM conveys another bits by the random combination of active indices. Thus, we demand two decision processes: a likelihood ratio test (LRT) detects $\mathbf{x}_l$ (and thus $m_0$); and the ML detector retrieves $m_1$ bits from the sub-carriers of the corrected and equalized active indices from the estimate $\hat{\mathbf{x}}_l$. This decoder is optimal at the cost of the additional decoder of sub-carrier index (e.g., see \cite{basar_tsp13} for details).

\subsection{Number of bit errors of the MCIK}

We examine $m_{e0,\beta}$. For this, let $(\mathbf{s}_{\beta,\alpha} \rightarrow \tilde{\mathbf{s}}_{\beta,{\tilde\alpha}})$ denote the pairwise error event (PEE) that in cluster~$\beta$ $\alpha$ is incorrectly detected as $\tilde\alpha$ for $\alpha,\tilde{\alpha} \in\{1,\cdots,N\}$ and $\alpha \neq \tilde{\alpha}$, given that $\alpha$ is transmitted within cluster~$\beta$. Then, given the PEE and $\mathbf{h}$, $m_{e,\beta}$ of each cluster can be obtained, using the union bound, as
\begin{equation}
m_{e,\beta} \leq \sum_\alpha \sum_{\tilde\alpha\neq \alpha} P\left( \mathbf{s}_{\beta,\alpha} \rightarrow \tilde{\mathbf{s}}_{\beta,{\tilde\alpha}} \right)
%P( \mathbf{s}_{\beta,\alpha} )\, \tilde m_{e,\beta}(\alpha, \tilde\alpha)
\frac{1}{N} \, \tilde m_{e,\beta}(\alpha, \tilde\alpha)
\label{eq:me}
\end{equation}
where $\tilde m_{e,\beta}(\alpha, \tilde\alpha)$ denotes the number of bit errors on both the sub-carrier index domain and the M-ary complex domain, i.e., $\tilde m_{e,\beta}(\alpha, \tilde\alpha)=\tilde m_{e0,\beta} + \tilde m_{e1,\beta}$. Here, $\tilde m_{e0,\beta}$ and $\tilde m_{e1,\beta}$ are for $m_{e0,\beta}$ and $m_{e1,\beta}$, respectively, conditioned on the PEE $(\mathbf{s}_{\beta,\alpha} \rightarrow \tilde{\mathbf{s}}_{\beta,\tilde\alpha})$. In \eqref{eq:me}, $P(\mathbf{s}_{\beta,\alpha} \rightarrow \tilde{\mathbf{s}}_{\beta,\tilde\alpha} )$ is the conditional pairwise error probability (PEP) of deciding $\tilde{\mathbf{s}}_{\beta,\tilde\alpha}$ given that $\mathbf{s}_{\beta,\alpha}$ and $\mathbf{h}$ are used, and the priori probabilities of $\mathbf{s}_{\beta,\alpha}$ are equally likely.

Using the LRT in \cite{smkay}, the well-known conditional PEP expression in \eqref{eq:me} can be expressed:
%, given $(\mathbf{s}_{\beta,\alpha} \rightarrow \tilde{\mathbf{s}}_{\beta,\tilde\alpha})$:
\begin{equation}
\label{eq:cpep}
P(\mathbf{s}_{\beta,\alpha} \rightarrow \tilde{\mathbf{s}}_{\beta,\tilde\alpha}) = Q\left( \frac{1}{2} \sqrt{\frac{|| h(i_\beta)\, (\mathbf{s}_{\beta,\alpha} - \tilde{\mathbf{s}}_{\beta,\tilde\alpha}) ||^2}{N_0}} \right)
\end{equation}
where $Q(x)\triangleq\pi^{-1}\int_{0}^{\pi/2}{\rm e}^{-x^2 /
2\sin^2\theta} \mathrm{d}\theta$ is the error function. Define $d_{\beta}= \gamma_\beta \| (\mathbf{s}_{\beta,\alpha} - \tilde{\mathbf{s}}_{\beta,\tilde\alpha}) \|^2= 2 E_s \gamma_\beta$, where $\gamma_\beta=|h(i_\beta)|^2$ has a chi-squared distribution with two degrees of freedom. Then, \eqref{eq:cpep} can be simplified to
\begin{equation}
\label{eq:cpep2}
P(\mathbf{s}_{\beta,\alpha} \rightarrow \tilde{\mathbf{s}}_{\beta,\tilde\alpha}) = Q\left( \sqrt{\frac{\gamma_\beta E_s}{2 \,N_0}} \right).
\end{equation}

Given \eqref{eq:cpep2}, $m_{e0,\beta}$ from \eqref{eq:me} can be obtained as
\begin{equation}
m_{e0,\beta} \leq \frac{1}{N} \sum_{\alpha=1}^N \sum_{\tilde\alpha\neq\alpha=1}^N Q\left( \sqrt{\frac{\gamma_\beta E_s}{2 \,N_0}} \right) \tilde m_{e0,\beta}.
\label{eq:meb}
\end{equation}

In \eqref{eq:meb}, $\tilde m_{e0,\beta}$ can be obtained referring to the number of unmatched active sub-carriers on the PEE in \eqref{eq:cpep2} by
\begin{equation}
\label{eq:tilde_me0}
\tilde m_{e0,\beta}= ||\mathbf{x}^+_\alpha - \mathbf{x}^+_{\tilde\alpha} ||^2   \qquad  \forall \alpha,\tilde\alpha
\end{equation}
where $\mathbf{x}^+_a$ for $a\in\{\alpha,\tilde\alpha\}$ is an index-to-binary mapping so that the index $a$ represents a sequence of $(\log_2 N)$ bits, meaning that \eqref{eq:tilde_me0} is the Hamming distance between $\alpha$ and $\tilde\alpha$ (denoted by $H(\alpha,\tilde\alpha)$) that counts the number of bit errors caused by incorrect indices of sub-carriers.

Using \eqref{eq:tilde_me0} and \eqref{eq:meb}, $m_{e0,\beta}$ as a part of $m_{e,\beta}$ in \eqref{eq:pb} can be obtained based on the MCIK per cluster by
\begin{equation}
m_{e0,\beta}=\frac{1}{N}\sum_{\alpha=1}^N \sum_{\tilde\alpha\neq \alpha=1}^N Q\left( \sqrt{\frac{d_\beta}{4 N_0}} \right) H(\alpha,\tilde\alpha).
\label{eq:me0}
\end{equation}
As seen in \eqref{eq:me0}, it is worth pointing out that $m_{e0,\beta}$ scales proportionally with the Hamming distance between binary sub-carrier indices while it decreases exponentially with the Euclidean distance $d_\beta=\gamma_\beta \| (\mathbf{s}_{\beta,\alpha} - \tilde{\mathbf{s}}_{\beta,\tilde\alpha}) \|^2$.

\subsection{Number of bit errors of the conditional OFDM}

We examine $m_{e1,\beta}$ of the M-QAM symbols. One term is added to \eqref{eq:me}, taking into account the bit errors of case~(iii). This is because the bit errors of the M-ary symbols can still occur even if the active sub-carriers are correctly detected.

So, $m_{e1,\beta}$ can be formulated for a given $\beta$ as
\begin{align}
\label{eq:cme1_1}
m_{e1,\beta} &\leq  \sum_\alpha \sum_{\tilde\alpha\neq \alpha} P\left( \alpha \rightarrow \tilde\alpha \right) \frac{1}{N} \, \tilde m_{e1,\beta}  \\
\label{eq:cme1_2}
&+ \sum_\alpha \left( 1 - \prod_{\tilde\alpha\neq \alpha} P\left( \alpha \rightarrow \tilde\alpha \right) \right) \frac{1}{N} \, \breve m_{e1,\beta}.
\end{align}
As shown in \eqref{eq:cme1_1}-\eqref{eq:cme1_2}, $m_{e1,\beta}$ relate the two terms to case~(ii) and case~(iii): (1) conditional bit errors (CBEs) on the mis-detection of the active indices; and (2) CBEs on the correct detection of the active indices.

\subsubsection{CBEs on the mis-detected active indices} This CBE has regard to $\tilde m_{e1,\beta}$ from $\tilde m_{e,\beta}(\alpha, \tilde\alpha)$ in \eqref{eq:me}. Given the PEE ($\alpha_\beta \rightarrow \tilde \alpha_\beta$) for $\alpha_\beta\neq \tilde\alpha_\beta$, $s(i_\beta)$ should be determined from a non information-carrying sub-carrier. It means that $\tilde m_{e1,\beta}$ is determined without any knowledge of $s(i_\beta)$, leading to $\tilde m_{e1,\beta}=0.5\log_2 M$.

Then, \eqref{eq:cme1_1} for the CBEs on the mis-detected indices can be captured for cluster~$\beta$ as
\begin{equation}
\label{eq:tilde_me1}
\frac{1}{N}\sum_{\alpha=1}^N \sum_{\tilde\alpha\neq \alpha=1}^N Q\left( \sqrt{\frac{d_\beta}{4 N_0}} \right) \frac{ \log_2 M }{2}
\end{equation}
where for a given $M$, $(\log_2 M) / 2$ represents $50$ percent detection accuracy of $\log_2 M$ transmit bits in the presence of the mis-detection, i.e., $\tilde\alpha \neq \alpha, \forall \alpha, \tilde\alpha$.

\subsubsection{CBEs on the correctly detected active indices}

We further derive \eqref{eq:cme1_2}. Intuitively, this equals the number of the bit errors in the classical M-QAM weighted by the probability of the correct detection that $\alpha \rightarrow \tilde\alpha$ for $\tilde\alpha=\alpha$. The probability of the correct detection of the active indices can be upper bounded by considering the joint probability of all PEEs. That is, \eqref{eq:cme1_2} for the CBEs can be represented by
\begin{equation}
\frac{1}{N}\sum_{\alpha=1}^N \left( 1- \prod_{\tilde\alpha\neq\alpha=1}^N Q\left( \sqrt{\frac{d_\beta}{4 N_0}} \right) \right) \log_2 M P(\gamma_\beta|s(i_\beta))
\label{eq:pcd}
\end{equation}
where the term $\left( 1 - \prod (\cdot) \right)$ including the product of the PEPs is used to give a upper bound on the correct detection probability of the active sub-carriers, and
$P(\gamma_\beta|s(i_\beta))$ stands for the well-known BER of the M-ary QAM over the AWGN channel. For example, given $s(i_\beta)\in {\cal S}$ and the M-QAM, we have \cite{chha03}
\begin{equation}
P(\gamma_\beta|s(i_\beta))=\sum_{i=1}^{\Theta_M}C_{i} Q\left( \sqrt{c_{i} \, \gamma_\beta \rho} \right)
\label{eq:ber_awgn}
\end{equation}
where for a Gray-coded square M-QAM, the
constants $\Theta_M$, $C_{i}$, and $c_{i}$ can be found in
\cite{chha03}.

Using \eqref{eq:tilde_me1}-\eqref{eq:ber_awgn}, therefore, $m_{e1,\beta}$ of the conditional OFDM on cases (ii)-(iii) per cluster can be given by
\begin{equation}
\begin{split}
&m_{e1,\beta} = \frac{\log_2 M}{N}\sum_{\alpha=1}^N \Bigg\{ \sum_{\tilde\alpha\neq\alpha=1}^N  \frac{1}{2} Q\left( \sqrt{\frac{d_\beta}{4 N_0}} \right)   \Bigg.\\
\quad &+ \Bigg. \left( 1 - \prod_{\tilde\alpha\neq\alpha=1}^N  Q\left( \sqrt{\frac{d_\beta}{4 N_0}} \right) \right) P(\gamma_\beta|s(i_\beta)) \Bigg\}
\end{split}
\label{eq:me1_2}
\end{equation}
where notice that the first and the second terms represent the CBEs of the mis-detection and the CBEs of the correct detection of active sub-carriers, respectively. As observed in \eqref{eq:me1_2}, $m_{e1,\beta}$ for a given $\beta$ relies on only the Euclidean distance $d_\beta$, unlike $m_{e0,\beta}$ in \eqref{eq:me0}.

\subsection{Unconditional BER expression in closed--form}
Using the above observations, the overall BER in \eqref{eq:pb} can be obtained with respect to $m_{e0,\beta}$ and $m_{e1,\beta}$ of all the clusters. Then, \eqref{eq:pb} can be represented by
\begin{equation}
P_b = \frac{m_e}{m_t} = \frac{ \sum_{\beta=1}^n (m_{e0,\beta} + m_{e1,\beta}) }{ \log_2 N^n + n \log_2 M }.
\label{eq:pb2}
\end{equation}

Inserting \eqref{eq:me0} and \eqref{eq:tilde_me1} into the numerator of \eqref{eq:pb2},
the \emph{conditional} BER on the mis-detection is expressed in closed--form for given $N,n,$ and $M$ as
\begin{equation}
\begin{split}
P_{b,c} &\leq  \sum_{\beta=1}^n \sum_{\alpha=1}^N \sum_{\tilde \alpha\neq\alpha=1}^N Q\left( \sqrt{\frac{d_\beta}{4 N_0}} \right) \frac{ H(\alpha,\tilde\alpha) }{m_t \, N} \\
 &+ \frac{\log_2 M}{m_t \, N} \sum_{\beta=1}^n \sum_{\alpha=1}^N \sum_{\tilde \alpha\neq\alpha=1}^N \frac{1}{2} \, Q\left( \sqrt{\frac{d_\beta}{4N_0}} \right).
\end{split}
\label{eq:cber1}
\end{equation}

The BER in \eqref{eq:cber1} is not the final BER but the conditional BER on the mis-detection cases (i)-(ii) only. It means that at favorable channels, \eqref{eq:cber1} does not address the case (iii) when $\alpha=\tilde\alpha$, relying on that the CBEs on the correct detection will get dominant in the BER.

Instead, inserting \eqref{eq:me0} and \eqref{eq:me1_2} into \eqref{eq:pb2}, the generalized expression for \emph{unconditional BER} of the MCIK-OFDM can be finally obtained in closed--form:
\begin{equation}
\begin{split}
&P_{b} \leq \frac{1}{m_t \, N} \sum_\beta^n \sum_\alpha^N \sum_{\tilde \alpha\neq\alpha}^N Q\left( \sqrt{\frac{d_\beta}{4 N_0}} \right) H(\alpha,\tilde\alpha) \\
 &+ \frac{\log_2 M}{m_t \, N} \sum_\beta^n \sum_\alpha^N
\Bigg\{ \sum_{\tilde\alpha\neq\alpha}^N  \frac{1}{2} Q\left( \sqrt{\frac{d_\beta}{4 N_0}} \right) +  \Bigg.\\
\quad & \Bigg. \left( 1 - \prod_{\tilde\alpha\neq\alpha}^N  Q\left( \sqrt{\frac{d_\beta}{4 N_0}} \right) \right)
{\sum_{i=1}^{\Theta_M}C_{i} Q\left( \sqrt{ \tilde\rho \gamma_\beta } \right) \Bigg\} }
%P(\gamma_\beta|s(i_\beta)) \Bigg\}
\end{split}
\label{eq:ber}
\end{equation}
where $\tilde\rho=c_{i}\, \rho$, the first and the second terms relate to the CBEs of the MCIK and the OFDM, respectively, on the mis-detection, and the last term represents the the CBEs of the OFDM based on the correct detection of the active sub-carrier indices. Note that this union bound based expression will be tight, as verified in Fig.~1.

\begin{figure}
\centering
\includegraphics[width=7cm]{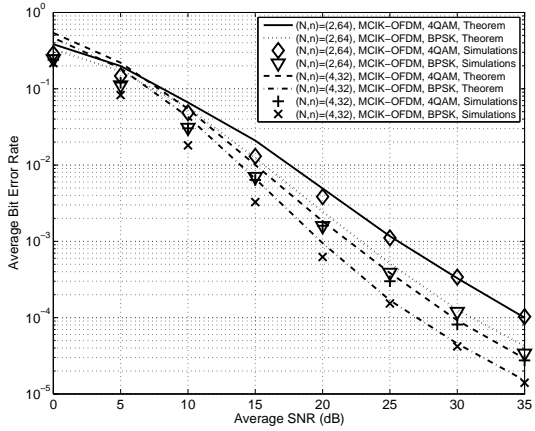}
\caption{Average BER performance of the MCIK-OFDM system in the presence of both imperfect and perfect detection of MCIK symbols over the Rayleigh fading per sub-carrier.}
\label{fig:fig1}
\end{figure}

%\begin{figure}
%\centering
%\includegraphics[width=7.3cm]{fig2_bpsk_qpsk_awgn.eps}
%\caption{Potential of the MCIK with a small $n=4$ experiencing both faster BER decrease and higher rate (e.g., 175\% increase) than the classical OFDM, when $M\in\{2,4\}$ and $(N,n)=(8,4)$.}
%\label{fig:fig2}
%\end{figure}

\section{Numerical evaluations and discussions}
\label{sec:num}

We consider the MCIK-OFDM systems with $N_c=128$ sub-carriers comprising of $n$ clusters of $N$ sub-carriers for various configurations of $(N,n)=\{(2,64), (4,32), (8,16)\}.$ The average BERs are obtained simply by taking the expectation of \eqref{eq:ber}.

Fig.~\ref{fig:fig1} depicts the average BER of the MCIK-OFDM on the Rayleigh flat fading per sub-carrier, considering the presence of both imperfect and perfect detection of active sub-carrier indices. The theoretical results are validated by simulations; the distance to the simulations decreases from 3~dB to within 1~dB as SNR increases. The accuracy improves further for the average BERs lower than $10^{-3}$.
This figure illustrates that the accuracy improves as $N$ (or $n$) decreases (increases). For small $N$, intuitively, the OFDM transmission gets a small number of the summation terms of the upper bound PEPs which improves the accuracy of the derived average BER.

\section{Conclusion}
We studied the MICK-OFDM system that modulates both the sub-carriers and their indices in order to convey the information bits via only a small subset of properly activated sub-carriers.
To measure the performance, we derived the tight upper bound BER expression in closed--form taking into account all the three conditional bit error cases on the activated index detection. The accuracy of the derived expression has been well validated by simulations and this accurate BER will be useful to evaluate various concepts of the MCIK-OFDM for low-complexity, energy-efficient applications.

%\section*{Acknowledgment}
%The research leading to these results has received funding from the European Union Seventh Framework Programme (FP7/2007-2013)
%under grant agreement number 316369 - project DUPLO.
%The authors would like to acknowledge
%the contributions of their colleagues from the DUPLO
%consortium.

\bibliographystyle{./IEEEtran}
\bibliography{./IEEEabrv,./ref_mcik_ko}

\newpage

%\begin{figure}%[hbt]
%\includegraphics[width=8cm]{fig4_cdfs.eps}
%\caption{The cumulative distribution function (CDF) of random power $|h_{ac,l}|^2$ after the analog SIC is depicted when $K=5,\rho_I=20~{\rm dB}, g_s=0~{\rm dB}$. For comparison, the CDFs of the desired data channel and the SI channel before the cancellation are illustrated.}
%\label{fig:fig4}
%\end{figure}

\end{document}